\documentclass[aps,prd,twocolumn,preprintnumbers,superscriptaddress,floatfix]{revtex4}

\usepackage{multirow, graphicx,amssymb,url,mathrsfs,amsmath}
\usepackage{eucal,wrapfig,boxedminipage,setspace,subfigure}
\usepackage{amsxtra,amstext,latexsym,dsfont}

\usepackage{xcolor}

                \def\g  {\gamma}

\def\IR{{\hbox{{\rm I}\kern-.2em\hbox{\rm R}}}}
\def\IB{{\hbox{{\rm I}\kern-.2em\hbox{\rm B}}}}
\def\IN{{\hbox{{\rm I}\kern-.2em\hbox{\rm N}}}}
\def\IC{\,\,{\hbox{{\rm I}\kern-.59em\hbox{\bf C}}}}
\def\IZ{{\hbox{{\rm Z}\kern-.4em\hbox{\rm Z}}}}
\def\IP{{\hbox{{\rm I}\kern-.2em\hbox{\rm P}}}}
\def\IH{{\hbox{{\rm I}\kern-.4em\hbox{\rm H}}}}
\def\ID{{\hbox{{\rm I}\kern-.2em\hbox{\rm D}}}}

\newcommand{\brac}[1]{\langle #1 \rangle}

\newcommand{\beq}{\begin{equation}}
\newcommand{\eeq}{\end{equation}}
\newcommand{\bea}{\begin{eqnarray}}
\newcommand{\eea}{\end{eqnarray}}

\begin{document}

\voffset 1cm

\newcommand\sect[1]{\emph{#1}---}

\title{Holography for QCD(Adj) and QCD(Adj)+F}

\author{Anja Alfano}
\affiliation{ STAG Research Centre \&  Physics and Astronomy, University of
Southampton, Southampton, SO17 1BJ, UK}

\author{Nick Evans}
\affiliation{ STAG Research Centre \&  Physics and Astronomy, University of
Southampton, Southampton, SO17 1BJ, UK}

\author{Wanxiang Fan}
\affiliation{ STAG Research Centre \&  Physics and Astronomy, University of
Southampton, Southampton, SO17 1BJ, UK}

\begin{abstract}
We discuss confinement and chiral symmetry breaking in SU($N_c$) gauge theories with fermions in the adjoint representation. There has been considerable work on studying these theories compactified on a small circle (with compactification scale $1/L$ large relative to the strong coupling scale of the theory). The weakly coupled IR theory of photons exhibits confinement through a density of magnetically charged instanton configurations. As the compactification scale $1/L$ approaches the strong coupling scale, the IR theory becomes strongly coupled. In this regime we propose a holographic description of the IR degrees of freedom. The instanton condensation scale can be associated with a scale at which the Brietenlohner-Freedman (BF) bound is violated in the model and the glueball spectrum computed. We can also introduce the adjoint fermions which holographically display a BF bound violation associated to their running anomalous dimension. Very naïvely extending the perturbative results to the non-perturbative regime suggests that chiral symmetry breaking might occur ahead of confinement, but equally they may be joined phenomena. If the two phenomena are separate, then it would be useful to be able to enlarge the gap in scales. We propose adding fermions in the fundamental representation as well, which in a holographic model (that favours this separation) can greatly enlarge the gap to an order of magnitude. These results challenge the lattice community to seek such scale gaps (or their absence) to further understand the confining and chiral symmetry breaking dynamics.
\end{abstract}

\maketitle

\newpage

\section{Introduction}\vspace{-0.5cm}

The fundamental mechanisms by which strongly coupled gauge theories, such as QCD, break chiral symmetries and confine remain to be clearly understood. The literature has split into two sectors. One group view chiral symmetry breaking as occurring due to the running coupling driving the anomalous dimension of the quark anti-quark operator through $\gamma=1$ (see for example \cite{Miransky:1984ef,Appelquist:1986an,Cohen:1988sq,Appelquist:1996dq,Ryttov:2007cx,Sannino:2009aw,Evans:2020ztq,Marciano:1980zf,Jarvinen:2011qe,Kutasov:2011fr,Alvares:2012kr} and the lattice papers measuring $\gamma$ with reference to this language such as \cite{LSD:2014nmn,Hasenfratz:2023wbr}). In this picture, confinement is a property of the low energy glue theory below the scale where the quarks have been integrated out. The alternative group have sought models where confinement is isolated and associated with non-perturbative monopole configurations
\cite{Seiberg:1994rs,Unsal:2007jx,Unsal:2007vu,Poppitz:2021cxe}. Here, as one moves back towards QCD, the monopole vacuum expectation value (vev) through effective Yukawa terms is responsible for chiral symmetry breaking. In QCD, these scales are presumably very close, and in reality it may be a mixture of these scenarios that occurs. It is interesting to continue to find theories in which the phenomena are separated. Here we will use holographic models of the two scenarios to contrast them and to look for separation mechanisms.

The ``$\gamma=1$'' construction was first motivated by Schwinger-Dyson truncation (gap equation) methods \cite{Miransky:1984ef,Appelquist:1986an,Cohen:1988sq,Appelquist:1996dq,Marciano:1980zf}. It is natural that an instability switches on when the running coupling grows enough to make the dimensions of the quark mass and condensate equal at $\Delta=2$. This picture has separately emerged in holography \cite{Jarvinen:2011qe,Kutasov:2011fr,Alvares:2012kr}, where the dimension of scalar operators is dual to the mass of a scalar in the AdS$_5$ bulk ($M^2=\Delta (\Delta -4)$)\cite{Witten:1998qj}. In rigorous top-down models, chiral symmetry breaking is triggered when the mass squared becomes radially dependent and passes in the IR through $M^2=-4/R$ (where $R$ is the AdS radius) - the Breitenlohner-Freedman instability bound \cite{Breitenlohner:1982jf}. Models in which supersymmetry and conformality are broken by a magnetic field or running coupling display this mechanism \cite{Babington:2003vm,Albash:2007bk}. The $\gamma=1$ criteria is often used to determine where the edge of the Conformal Window is for gauge theories as a function of the number of fermion flavours \cite{Appelquist:1996dq,Ryttov:2007cx,Sannino:2009aw}. Here it is believed that the Banks-Zaks fixed point, when $N_f$ lies close to the boundary for the loss of asymptotic freedom, gives way to a chiral symmetry breaking phase at lower $N_f$ where the IR fixed point coupling rises. Lattice studies of the conformal window often concentrate on the value of $\gamma$ to argue that a theory is IR conformal \cite{LSD:2014nmn,Hasenfratz:2023wbr,Rantaharju:2015cne,Athenodorou:2024rba,Bennett:2024qik}. Inherent in this world view is that instanton or monopole configurations appear in the deep IR below the chiral symmetry breaking scale.

On the other hand, there have been some very impressive pieces of work in which the confinement mechanism has taken centre stage \cite{Seiberg:1994rs,Unsal:2007jx,Unsal:2007vu,Poppitz:2021cxe}. The Seiberg-Witten theory of ${\cal N}=2$ supersymmetric Yang Mills (SYM) theory \cite{Seiberg:1994rs} is an example where chiral symmetry breaking is forbidden by the supersymmetry, and non-perturbative monopole configurations manifest. Breaking to ${\cal N}=1$ supersymmetry leads to their condensation and confinement. The gauginos acquire a mass via a Yukawa term to the monopole vev. Another impressive construction is the work in compactified  SU($N_c$) theories with fermions in the adjoint representation (QCD(Adj)) \cite{Unsal:2007jx,Unsal:2007vu,Poppitz:2021cxe}. When this theory is compactified on a small circle (relative to the inverse strong coupling scale), confinement by instanton configurations, which appear as a monopole density in the 3 dimensional IR theory, occurs at weak coupling. It is natural to speculate from these cases that non-perturbative configurations are potentially key to both confinement and chiral symmetry breaking.

In this paper, we wish to make a first step in bringing these pictures together. We will use holographic model building to attempt to unify the ideas. In particular, the condensation of monopoles and quarks must each be driven by a BF bound violation in the theory. Here we will build a very simple model of QCD(Adj) on a compact circle whose radius is close to the inverse strong coupling scale. The IR theory is expected to be strongly coupled and holography may be a sensible tool to study the spectrum. We will produce an AdS/QCD style model that describes the monopole condensate as forming due to a BF bound violation and the resulting bound state spectrum. Here we will just model theories with massive fermions (although below the strong coupling scale).

The model allows us to estimate the scale at which the instability sets in by extrapolating from the perturbative regime. Of course, this is very naïve. The perturbative exponential suppression of instanton configurations hints that this scale is somewhat lower than the equivalent one for chiral condensation (from looking at where $\gamma=1$ from one gluon exchange). However, the theory with a single massless Weyl fermion in the adjoint representation is ${\cal N}=1$ SYM, where the gaugino condensate and the glueballs are bound into a single supermultiplet \cite{Veneziano:1982ah}. This theory is likely an example where the mechanisms maximally converge, so whether they can be separated in other cases is unclear.

To play Devil's advocate it is interesting to try to conceive of theories where chiral symmetry breaking is separated from confinement. We have argued previously \cite{Evans:2020ztq}, that this might be supported in the $\gamma=1$ paradigm (see also \cite{,Marciano:1980zf}). In theories with fermions in multiple representations, higher dimension representations than the fundamental typically couple more strongly to gluons, and reach the $\gamma=1$ criteria ahead of the fundamental representation. If they condense and are integrated from the theory, then a gauge theory with just fundamentals is left in the IR, which presumably behaves like QCD (for sufficiently low $N_f^F$ flavours). If one can include sufficient fundamentals to slow the running between the higher representation condensing and the fundamental condensing, then presumably confinement is also separated. One needs to be careful however, not to push the theory into the conformal window by adding too much matter.

In a previous paper \cite{Alfano:2024aek}, we built a holographic model of SU($N_c$) theories with two-index symmetric matter and fundamentals which displayed such gaps. The model has no confinement mechanism because it is assumed to happen at or below the chiral symmetry breaking scale of the fundamental matter. Here we use the same model for SU($N_c$) with a single Weyl adjoint fermion and $N_f^F$ fundamentals - we pick a single Weyl adjoint to allow the maximum additional number of fundamentals to slow the running below the adjoint IR mass scale. We observe mass gaps for some choices of $N_c$ and $N_f^F$ as large as an order of magnitude. The gap size does depend on the extrapolations used for the running of the anomalous dimensions from the perturbative to non-perturbative regime as we investigated in \cite{Alfano:2024aek}. We also neglect interactions between the two fermionic sectors (and potentially the instanton/monopole sector) - condensation in one could trigger condensation in the other, for example, undoing the conclusions. It would be interesting in the future to study such mutual interactions. For the moment, the model presented is intended as a challenge to lattice studies to seek such phenomena.

The current state of lattice simulations for gauge theories with adjoint matter is as follows. The SU(2) gauge theories with both $N_f^{Adj}=1$ and 2 appear to lie in the conformal window (i.e. they do not break chiral symmetry) with anomalous dimensions around $0.2-0.3$ \cite{Rantaharju:2015cne,Athenodorou:2024rba,Bennett:2024qik}. ${\cal N}=1$ super Yang-Mills theory with $N_c=2$ and $N_c=3$ is known to break chiral symmetry and confine
\cite{Demmouche:2010sf,Ali:2018dnd,Ali:2019agk,Steinhauser:2020zth}. There has been one initial study of an SU(2) gauge theory with a single Weyl fermion and two Dirac fundamental fermions \cite{Bergner:2020mwl}, which indeed shows a gap between the $\rho$ mesons made of adjoint and fundamental matter (the adjoint $\rho$ is reported as being about 1.6 the mass of the fundamental $\rho$ - see Table 3 in \cite{Bergner:2020mwl}) and is consistent with our holographic model using the $\gamma=1$ criteria. A key future test of the separation of the confinement and chiral symmetry breaking mechanisms would be to check if they occur at different thermal transitions.

\vspace{-0.6cm}

\section{The Gauge Theories} \vspace{-0.25cm}

We will consider $SU(N_c)$ gauge theories with fermionic matter in the adjoint and, later, the fundamental representations. The two loop running of the gauge coupling in QCD for arbitrary representation is given by
\begin{align}
&\mu \frac{d\alpha}{d\mu} = -b_0 \alpha^2-b_1 \alpha^3\\
&b_0 = \frac{1}{2\pi} \left( \frac{11}{3} C_2(G) - \frac{4}{3} \sum_{R} T(R) N_f(R) \right). \\
&\nonumber b_1 = \frac{1}{24\pi^2} \left( 34 C_2^2(G) \right. \\
&  \left. - \sum_{R} \left[ 20 C_2(G) + 12 C_2(R)\right] T(R) N_f(R)  \right).
\end{align}
where $N_f(R)$ is the number of Dirac fermion flavours in the representation $R$. We will use the notation $N_f^F$ for the number of fundamental fermions and $N_f^{Adj}$ for the number of adjoint fermions. For $SU(N_c)$ theories $C_2(G)=N_c$; $T(Adj)=N_c$ and $T(F)=1/2$.
N.b. for Weyl fermion we must include a factor of 1/2 in the flavour count in these formulae.

To extract the running we must choose initial conditions for the RG equation. For example, $\alpha(Log(\Lambda_{UV}=5))=0.1393$, for $N_c=2$, $N_f=0$, gives a Landau pole at $Log(\Lambda)=0; \Lambda=1$. A more accurate way to set the strong coupling scale, is to set a bound state mass (we will take the $\rho$ meson made from the adjoint fermions) in any theory at zero fermion mass to be the strong coupling scale. We therefore, when comparing the spectra of theories, write all masses and couplings in units of the adjoint $m_\rho$ at $m_f=0$. 

The two-loop ansatz for the running includes IR fixed point behaviour - the so-called conformal window \cite{Appelquist:1996dq,Ryttov:2007cx,Sannino:2009aw}. As $N_f$ is lowered, the IR fixed point coupling grows and at some point is expected to trigger chiral symmetry breaking and or confinement - this is the edge of the conformal window. We will use our holographic models to estimate these critical couplings below. When we predict the spectra of the Weyl adjoint plus fundamentals theory, we will show plots over a range of $N_f^F$ up to the edge of the conformal window (from below) - the position of the edge in each case can therefore be seen on those plots to come.

\section{Confinement in Compact SU(2)(Adj) Gauge Theory}

For simplicity, we will initially restrict our discussion to the case of $N_c=2$ with adjoint fermions. Here, the review \cite{Poppitz:2021cxe} is very useful. We will first review the results for confinement in the theory on a small compact direction \cite{Unsal:2007jx,Unsal:2007vu,Poppitz:2021cxe}; then we will write a holographic model for the intermediate regime where the IR theory is strongly coupled; we finally briefly discuss the extension to SU($N_c$). We will then be able to consider  the interplay between chiral symmetry breaking and confinement in section IV.

\subsection{Summary of the SU(2)(Adj) theory on $R^3 \times S^1$}

There has been considerable work \cite{Unsal:2007jx,Unsal:2007vu,Poppitz:2021cxe} on understanding four dimensional SU(2) gauge theory with $N_f$ Weyl fermions in the adjoint representation on a compact circle of radius $L$. At the scale $1/L$ we can rewrite the 4d gauge field as a 3d gauge field and a real, adjoint scalar, $a_4$. The classical potential for the scalar, inherited from the 4d $Tr F^2$ commutator term, allows a vev that breaks SU(2) to U(1) and leaves a massless U(1) gauge field and a massless, chargeless scalar. One expects the vev to lie at the scale $1/L$ when loop corrections are included. Any charged adjoint fermions speak to the scalar vev through a Yukawa term (generated from the 4d covariant derivative) and are massive. The charge zero fermions survive in the IR theory. The (naïvely) non-interacting IR theory has a characteristic coupling $g_4^2(1/L)$. 

The consistency of this picture can be checked by computing the Coleman-Weinberg effective potential from the gauge and fermion fields (including their KK towers) and confirming that it is minimized at the vev $2 \pi/L$.  The potential for the vev $v$ is given by
\begin{equation}
    V = -\frac{(N_f-1)}{L^3}\frac{1}{12 \pi^2} [vL]^2 (2 \pi - [vL])^2, 
\end{equation}
where $[vL] = vL ({\rm mod}2 \pi)$. Note $v$ is the vev of $A_4^3$, and $a_4$ is the fluctuation about the vev, i.e. $A_4^3=v+ a_4$

For $N_f^{Adj}=1/2$ (SYM) the potential vanishes and one must argue that non-perturbative effects will stabilize the vev. For $N_f>1$, the minimum is at $v=2\pi/L$. Note that the fluctuation of the scalar about the minimum has a mass of order $g/L$ and the $a_4$ can thus be integrated from the IR theory. In fact even if the adjoint fermions have masses $m \leq 1/L$ they still act to stabilize the vev and can then be neglected in the deep IR - we will work in this theory here for simplicity.

One can use electromagnetic duality in 3d to rewrite the IR U(1) electric and magnetic fields in terms of a single (dimensionless) scalar potential $\sigma$:
\begin{equation}
    F^{\mu \nu} = \left( \begin{array}{ccc} 0 & E_x & E_y \\ -E_x & 0 & B \\ -E_y & -B & 0 \end{array} \right) \leftrightarrow \partial_\mu \sigma = \frac{4 \pi L}{g_4^2} ~( -B, E_y, -E_x).
\end{equation}
The kinetic term for $\sigma$ is
\begin{equation}  \label{norm}
{\cal L_\sigma}= \frac{1}{2} \frac{g_4^2}{(4 \pi)^2 L} ~ (\partial_\mu \sigma)^2.
\end{equation}

The interesting aspect of the theory is that SU(2) instanton dynamics above the breaking scale generate confinement in the low energy theory. The mechanism is the 3d version of the dual Meissner effect. In particular, note that the natural charges for a 0-form potential are pseudo-particles (as a vector couples to point particles and a two-index field to strings, etc.) So, for example, a constant density of magnetically charged pseudoparticles at some time $t=0$ emits field lines into the time direction in analogy to a charged plane emitting field lines in the perpendicular spatial direction in 3+1d. The solution, using Gauss' law, is $\sigma = B t$ and there is a constant magnetic field in the space. This motivates the idea that the instantons of the SU(2) theory are suitable candidates to play the role of such magnetic charges. Indeed, explicit construction shows that the instantons do indeed radiate magnetic field lines asymptotically. These computations have been done in detail \cite{Lee:1997vp,Kraan:1998pm} - there are two types of instantons: M and KK instantons, which are both magnetically charged.

We would expect the instanton and anti-instanton distributions to be uniform on average at all $x,y,t$, so there is no net magnetic field from their presence. They do though, generate a potential for $\sigma$. If we stick to the theory where the fermions have a small mass, then a potential is generated directly by the M and KK instantons. It is given by
\begin{equation} \label{inter}
    V = \frac{4 e^{-S_0}}{L^3} (1 - \cos \sigma) 
\end{equation}
where $S_0= \frac{4 \pi^2}{g_4^2(1/L)} $ is the action of the instanton configuration, which is minimized at $\sigma=0$ and the $\sigma$ field has an effective mass \[ m_\sigma^2 = \frac{64 \pi^2 e^{-S_0}}{g_4^2 L^2} \]

Note, the theories with massless fermions are more complicated since they possess a remnant of the anomalous $U(1)_A$ symmetry that acts as a $Z_{4N_f}$, under which $\sigma$ shifts by $\pi$. This shows that the M and KK contributions to the potential must vanish in these theories. It is then possible to consider dyons made of bound M and KK states and argue that magnetic charge two dyons play the important role of condensing and causing confinement \cite{Unsal:2007jx}. Let us stay in the massive theory for simplicity though. 

Confinement can be seen directly because there are excited states of the vacuum that correspond to electric flux tubes. Here, one finds solutions for $\sigma$ that traverse from 0 to $2 \pi$ as one moves across a line in the  $x-y$ plane - for example, in $x$ in a $y$-independent solution. The solution of the classical equation of motion for $\sigma$ is
\begin{equation}
    \sigma(x) = 4 \arctan e^{-m_q x}
\end{equation}
Asymptotically at large $x$, this solution's action returns to that of the vacuum. In the central region, the solution lies at the top of the potential where the $\sigma$ mass falls to zero (here the instanton density has fallen to zero) and $\partial_x \sigma = E_y \neq 0$. The flux tube's energy will be proportional to the length of the tube in $y$ (neglecting end effects where the electric charges attach).
\vspace{-0.5cm}

\subsection{From Weak to Strong Coupling with Holography} \vspace{-0.5cm}

One of the strengths of the work performed on the adjoint theory is that when the compact radius is small - so $1/L$ is large compared to the strong coupling scale of the field theory - one can see the confinement dynamics at weak coupling. Nevertheless, it is interesting to consider the transition to strong coupling. As the compactification length rises and the scale $1/L$ falls, eventually $g_4(1/L)$ will become large and perturbation theory will break down. In this regime, one can propose a holographic description of the strongly coupled $\sigma$-instanton bath system. This description will hold until the field vevs grow to $1/L$, when one should return to a purely 4d description.

The natural starting point is to place the effective IR action into AdS$_4$ space 
\begin{equation}
    ds^2 = {\rho^2 dx_{2+1}^2 + \frac{d \rho^2}{\rho^2} }
\end{equation}
(here $x_{2+1}$ are the Minkowski directions of the field theory, and $\rho$ is the radial direction that becomes the renormalization group scale). We include a dimension 3 field $I$, that will correspond to the instanton density, and a dimensionless field $\hat{\sigma}$ which is the holographic partner to the field theory $\sigma$ operator. 

Our action is
\begin{align}\label{eq: dim 3 action}
    S^{AdS_4}=& \int d\rho d^3 x ~~
    \frac{1}{2}\frac{\rho^3}{r^2} \frac{\cal C}{L}G^{MN} \partial_{M}\hat{\sigma} \partial_{N}\hat{\sigma} +\frac{I}{r}2 \sin^2{\frac{\hat{\sigma}}{2}}\nonumber
    \\ & +\frac{1}{2}\left(\frac{1}{\rho^2} \frac{1}{r^2}G^{MN} \partial_{M} I\partial_{N} I
    +M_{I}^2 \frac{I^2}{r^4}\right)
\end{align}
Note here we use symbols $r$ and $\rho$, which for the moment are the same (we will distinguish them when we come to discuss fluctuations below). 
The constant ${\cal C}=\frac{g_4^2}{(4\pi)^2 }$ is taken from the field theory action normalization (\ref{norm}). The field $\hat{\sigma}$, assuming $I \rightarrow 0$ in the UV, has UV solution
\begin{equation}
\frac{1}{L} \partial_\rho ( {\rho^3} \partial_\rho \hat{\sigma}) =0, \hspace{1cm}    \hat{\sigma} = c + \frac{c'}{\rho^2}
\end{equation}
$\hat{\sigma}$ is dimensionless. We interpret $c$ as the dimensionless field vev and $c'/L$ as the dimension 3 source for the field. 

$M_I^2$ is a mass term for $I$, which if zero in the UV (and $\hat{\sigma} \rightarrow 0$ in the UV), ensures the UV equation of motion and solution for $I$ is
\begin{equation}
\partial_\rho ( \frac{1}{\rho^2} \partial_\rho I) =0, \hspace{1cm}I = k + k' \rho^3
\end{equation}
$I$ has dimension 3. Here we interpret $k$ as the vev of the dimension 3 field and $k'$ as its dimensionless source. We will pick an example form for $M_I^2$ shortly.

The interaction term is that in (\ref{inter}), where we need the field $I$ to acquire a vev $4 E^{-S_0}/L^3$. To enforce that, consider the action that controls the vacuum, where we assume that the fields do not depend on the spatial or time directions but only on $\rho$. The bulk Lagrangian density reduces to 
\begin{align} \label{l4}
\mathcal{L}^{AdS_4}= &\frac{1}{2} \rho^3 \frac{{\cal C}}{L}(\partial_{\rho}\hat{\sigma})^2+ \frac{1}{2} \left( \frac{1}{\rho^2} (\partial_{\rho} I)^2 +M^2_{I} \frac{I^2}{r^4} \right) \nonumber\\&+ \frac{I}{r} 2 (\sin{\frac{\hat{\sigma}}{2}})^2
\end{align}

In fact, the interaction term pins $\hat{\sigma}=0$ in the vacuum solutions with non-zero $I$. We must solve
\begin{equation} \label{vac}
    \partial_\rho ( \frac{1}{\rho^2} \partial_\rho I) - M_I^2 \frac{I}{\rho^4} - \frac{I^2}{2 \rho^4} \frac{\partial M_I^2}{\partial I} = 0
\end{equation}

It is necessary to choose IR and UV boundary conditions. Here we allow ourselves to be led by the D3/D7 probe system \cite{Karch:2002sh,Kruczenski:2003be,Erdmenger:2007cm}, where the D7 brane embedding solutions for the massless theory can be interpreted as the source vanishing in the UV and the operator vanishing in the IR. (The DBI action for the D7 brane embedding field $\chi$, takes the form ${\cal L} = \rho^3 (\partial_\rho \chi)^2$, and the solution is of the form $\chi= m + c/\rho^2$, where $m$ is the quark mass and $c$ the quark condensate operator. Here, one seeks solutions where $m$ vanishes in the UV for the massless theory, and in the IR, a regular brane embedding requires $\chi'_{IR}=0$, i.e. vanishing condensate.)

Thus for $I$ we solve using
\begin{equation}
    I(\rho_{IR}) = \rho_{IR}^3, \hspace{1cm} \partial_\rho I|_{\rho_IR} = 3 \rho_{IR}^2
\end{equation}
The first condition is an on mass-shell condition for the field below which it should be integrated from the dynamics. The second condition ensures that the solution tends to a constant, dynamically generated source value at the on mass-shell point.

For $\hat{\sigma}$ and its fluctuations we will use the - similarly justified - IR boundary conditions
\begin{equation}
    \hat{\sigma}(\rho_{IR}) = 1/\rho_{IR}^2, \hspace{1cm} \partial_\rho \hat{\sigma}|_{\rho_IR} = -2 /\rho_{IR}^3
\end{equation}
and require in the UV
\begin{equation}
    \rho^3 \partial_\rho \hat{\sigma} = 0 
\end{equation}
Now we must decide on a form for the mass squared for $I$. We want a running mass (i.e. a $\rho$ dependent mass) that will violate the BF bound in the IR causing condensation of $I$ - we will adjust the BF bound violation point so that the condensation occurs to match that expected in the field theory model. Note that with the chosen dimension for I, $M_I^2=-9/4$ is the BF bound violation point where $\Delta = 3/2$. A simple choice we can make is
\begin{equation} \label{MI}
    M_I^2 = -\frac{{\cal K}}{r}
\end{equation}
at low $r$ (for the moment $\rho$), this will violate the BF bound. If one were to set purely $r=\rho$, then there would be no stable IR solution since the BF bound is violated for all $I$ vevs. A simple resolution, that occurs naturally in the D3/probe D7 system \cite{Karch:2002sh,Kruczenski:2003be,Erdmenger:2007cm} for example, and which is dimensionally consistent is to use
\begin{equation}
    r^2 = \rho^2 + I^{2/3}.
\end{equation}

\begin{center}
\includegraphics[width=7cm,height=5cm]{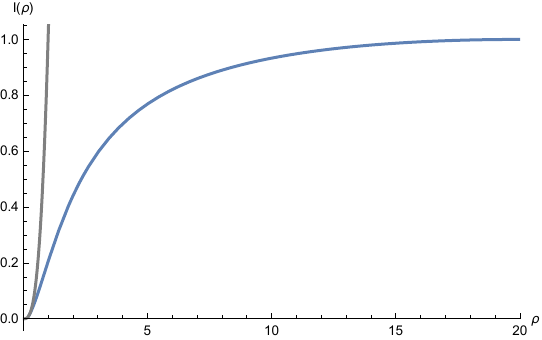}
\includegraphics[width=7cm,height=5cm]{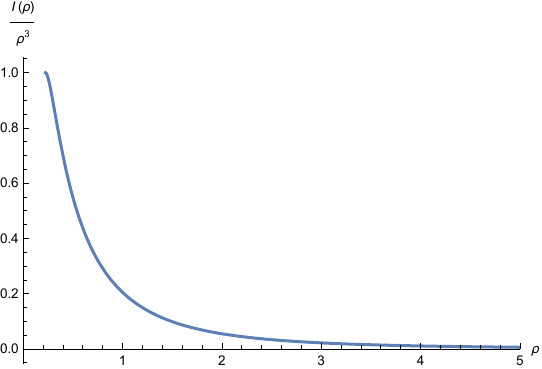}\\
\textit{Figure 1: Top: The blue line is the numerical result for the vacuum solution $I(\rho)$. The grey line represents the function $y=x^3$ which sets the IR boundary. Bottom: The blue line shows the plot $\frac{I(\rho)}{\rho^3}$ against ${\rho}$. (Set-up: ${\cal K}=7.82498$, $\rho_{IR}=0.223527$, $\rho_{UV}=20$ such that $I(\rho_{UV})=1$ and $I'(\rho_{UV})=0$.) }
\end{center}

Now, if $I$ acquires a vev, it can move to a value where the BF bound is not violated and it becomes stable. At this point, we also replace occurrences of $r$ in the action with $\rho$ as shown in (\ref{l4}) - here, again as in the D3/probe D7 system, the replacements feed the presence of the vev to the fluctuations, but are introduced so as not to change the UV asymptotic solutions. Although this choice looks a little arbitrary, the D3/probe D7 system artfully enables this from first principles - we follow it's example.

The constant ${\cal K}$ is the first introduction of a scale into the action for $I$ ($L$ does not enter when $\hat{\sigma}=0$). Numerically, we have solved (\ref{vac}) with (\ref{MI}) to find the value of ${\cal K}$ so that the solution in the UV tends to $I=1$ with vanishing source (derivative). We find ${\cal K}=7.82498, \rho_{IR}=0.223527$. We plot the solution in Figure 1 (and also the form of $I/ \rho^3$ which displays the solution as a running source term more analogous to the familiar D7 probe embeddings in the D3/D7 system).  

Next, we consider the fluctuation around the vacuum solutions: $\bar{I}(\rho)+\epsilon I(\rho, x) $ and $\hat{\sigma}(\rho)=0+\epsilon \hat{\sigma}(\rho,x)$, expanding the action \eqref{eq: dim 3 action}  to $\mathcal{O}(\epsilon^2)$. We seek perturbations of the form $I(\rho, x)=f_1(\rho) e^{i k_1 \cdot x}$ and $\hat{\sigma}(\rho, x)=f_2(\rho) e^{i k_2 \cdot x}$, with $k_1^2=-M_1^2$ and $k_2^2=-M_2^2$ (those being the masses of the bound states). With some algebra, the equations of

\begin{center} \label{fig:dim3_inst_fluc}\includegraphics[width=7cm,height=5cm]{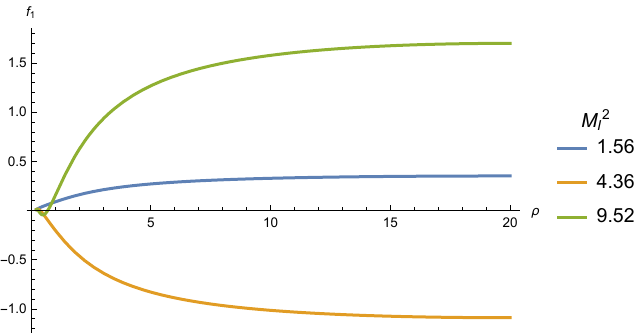}
\includegraphics[width=7cm,height=5cm]{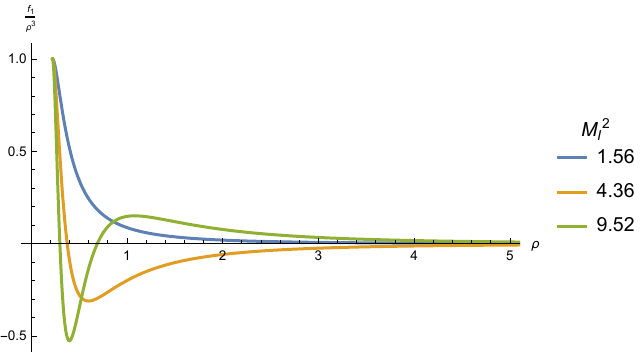}\\
\textit{Fig 2: Top: The regular solutions for the fluctuation of the instanton density $f_1$ against $\rho$. Bottom: $\frac{f_1}{\rho^3}$ against $\rho$.}
\end{center}

 motion for the fluctuations are
\begin{align}\label{eom:sigma_f2}
    \frac{\bar{I}f_2(\rho)}{\sqrt{\bar{I}^{2/3}+\rho^2}}-\partial_{\rho}\left(\frac{{\cal C}}{L}\rho^3 \partial_{\rho}f_{2}(\rho)\right)\nonumber\\-\frac{\rho^3}{(\rho^2+\bar{I}^{2/3})^2} \frac{\cal C}{L}M_2^2 f_2(\rho)=0
\end{align}
\begin{align}\label{eom:I_f1}
    -\frac{k \left(-19 \rho ^2 \bar{I}^{2/3}-2 \bar{I}^{4/3} + 18 \rho ^4\right)}{18 \left(\bar{I}^{2/3}+\rho ^2\right)^{9/2}}f_1(\rho)\nonumber\\-\partial_{\rho}\left(\frac{1}{\rho^2}\partial_{\rho}f_1(\rho)\right)-\frac{M_1^2}{\rho^2(\rho^2+\bar{I}^{2/3})^2}f_1(\rho)=0,
\end{align}
which implies that the fluctuations $f_1,f_2 $ decouple. And the coefficient ${\cal C}/L=\frac{g_4^2}{L(4\pi)^2}$ will affect the spectrum of the dual photon.

The equations of motion for the instanton fluctuation are solved with boundary conditions $f_1(\rho_{IR})=\rho_{IR}^3$, $f_1'(\rho_{IR})=3 \rho_{IR}^2$. The meson masses are obtained by fine-tuning $M_1^2$ so that $f_1'(\rho_{UV})\rightarrow0$. The mass spectrum, in the theory with $I \rightarrow 1$ in the UV, for the instanton density fluctuation is $M_I^2 = 1.56, 4.36, 9.52$, with corresponding excitation number $n=0, 1, 2$, and the corresponding solutions are presented in Fig 2.

Next, we compute the meson modes for $\hat{\sigma}$. The boundary conditions used are $f_2(\rho_{IR})=\frac{1}{\rho_{IR}^2}, f_2'(\rho_{IR})=\frac{-2}{\rho_{IR}^3}$, and 

\begin{center} \label{fig:cL_g4}\includegraphics[width=7cm,height=5cm]{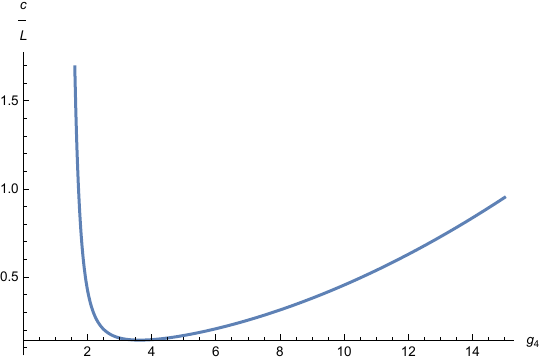}
\includegraphics[width=7cm,height=5cm]{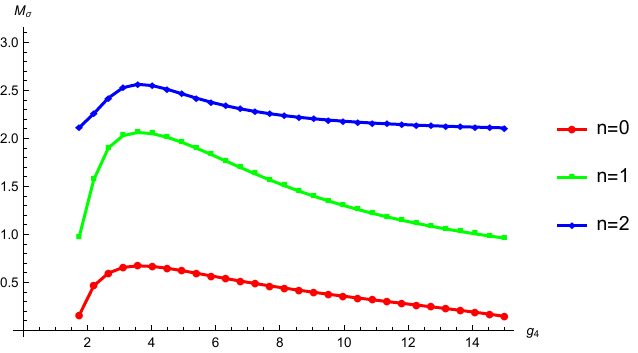}\\
 \textit{Fig 3: Top: $\frac{c}{L}$ against $g_4$ in Eq. \eqref{cl_g4}; There is a minimum ${c/L}=0.143$ at $g_4=\frac{2\pi}{\sqrt{3}}=3.7$. Bottom: $M_\sigma$ against $g_4$.}\label{fig_3}
\end{center}

$\rho_{UV}^3 f_2'(\rho_{UV})=0$, which make the condensate vanish in the IR, and the source vanish in the UV. The results depend on the value of ${\cal C}/L$. Here we rewrite that scale in terms of the $I$ vev as
\begin{align}
    &\langle I \rangle = \frac{4 e^{-S_0}}{L^3}, \hspace{0.3cm}
    S_0=\frac{4 \pi^2}{g_4^2}\\
    &\frac{{\cal C}}{L}=\frac{g_4^2}{(4 \pi)^2 L}=\sqrt[3]{\frac{\langle I \rangle}{4 e^{-S_0}} }\,\frac{g_4^2}{(4\pi)^2}\label{cl_g4}
\end{align}
Now we compute the $\hat{\sigma}$ spectrum as a function of $g_4$ which directly controls ${\cal C}/L$ at $\langle I \rangle = 1$. The masses can be directly compared to those in Fig 2 for the $I$ fluctuations. This suppresses the $I$ dependence on $L$, or equivalently $g_4$. The $\hat{\sigma}$ masses are plotted in Fig 3.

It is worth noting that when $c/L$ is large (when $g_4<1.72$ or $g_4>15$), there is a tachyon mode, i.e, $m_{\sigma}^2<0$. This follows from computing the spectrum using (\ref{eom:sigma_f2}) neglecting the first term when $M_\sigma^2=-0.298$. These regimes correspond to parameter choices where the holographic 
model should not be applied. When the gauge theory is weakly coupled, the holographic dual should become strongly coupled. Equally, when the theory has $g_4 > 4 \pi$, the four dimensional theory will hit strong coupling before reaching the IR compactification scale and there should never be a three dimensional description. It is interesting that the holographic model becomes aware of these regimes where it fails. 

\begin{center} \label{fig:mass spectrum sigma_1 and 2}\includegraphics[width=7cm,height=5cm]{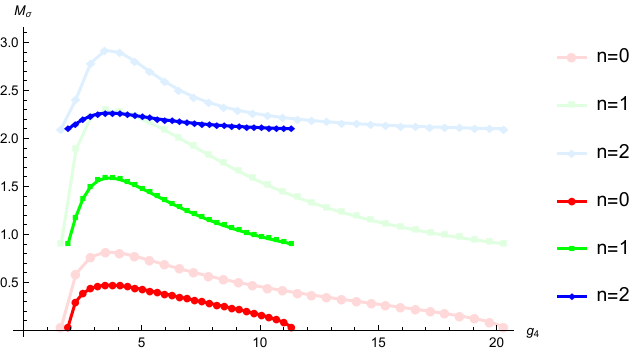}\\
\textit{Fig 4: The mass spectrum of dual photons $\sigma_1$ and $\sigma_2$  for SU(3). The darker colours represent $\sigma_1$; The lighter colours represent $\sigma_2$.}
\end{center}

\subsection{The SU($N_c$) Theory} 

The SU($N_c$) theory with adjoint matter (which we again assume is massive, but with those masses below the compactification scale) shows similar behaviour \cite{Unsal:2007jx,Unsal:2007vu}. The $A_3$ component of the gauge field again becomes an adjoint scalar on compactification and its vev breaks the theory to U(1)$^{N_c-1}$ - there are $\sigma_i$ fields ($i=0..N_c-2$). The instanton monopoles are now ``bi-fundamental'' fields with charge $(+1,-1)$ under the adjacent U(1)s one would obtain from breaking U($N_c$) (one can just switch off the coupling of the extra U(1) to reduce to SU($N_c$)). 

For example, for SU(3) the resulting potential for the two $\sigma_i$ are
\begin{equation}
    V =  \frac{4 e^{-S_0}}{L^3} \left( \sin^2 \frac{\sigma_0}{2} + \sin^2 \frac{\sigma_1}{2} +  \sin^2\frac{\sigma_0 - \sigma_1}{2} \right)
\end{equation}
with $S_0\rightarrow 8 \pi^2/ g_4^2 N_c$.

At the level of the holographic model, one includes three copies of the monopole field $I$ and an appropriate potential. The holographic model for SU(3) is
\begin{align}
    S^{AdS_4}&= \int d\rho~ d^3 x ~~
    \frac{1}{2}\frac{\rho^3}{r^2} \frac{\cal C}{L}G^{MN} \partial_{M}\sigma_i \partial_{N}\sigma_i \nonumber\\
     & +\frac{1}{2}\left(\frac{1}{\rho^2} \frac{1}{r^2}G^{MN} \partial_{M} I_i\partial_{N} I_i
    +M_{I}^2 \frac{I_i^2}{r^4}\right) \\
    &+\frac{I_1}{r}\sin^2(\frac{\hat{\sigma}_1}{2})+\frac{I_2}{r}\sin(\frac{\hat{\sigma}_2}{2})+\frac{I_3}{r} \sin(\frac{\hat{\sigma}_1-\hat{\sigma}_2}{2})\nonumber
\end{align}
where we chose the $M_I^2$ so that   $I=4 e^{-S_0}/L^3$, and ${\cal C}=(\frac{g}{4\pi})^2 $. The vacuum solutions for the I fields lead to the same vev solutions, and there is a multiplicity of the same excitation states as we saw in SU(2). 
For the $\hat{\sigma}_i$ fluctuations the potential can be daigonalized by writing
\begin{align}
    \tilde{\sigma}_{\pm} =\frac{1}{\sqrt{2}}(\hat{\sigma}_1\pm\hat{\sigma}_2)
\end{align}
This leads to two mass eigenstates ($\hat{\sigma}_0 \pm \hat{\sigma}_1$). We show the resultant mass spectrum in Figure 4.

\section{Comparison of Instabilities}

The BF bound violating scale \cite{Breitenlohner:1982jf} marks the onset of the instability to condensation of an operator in the holographic context.

In the SU(2) model where the instanton density vev was 1 we needed ${\cal K}=7.82498$.  The BF bound violation  occurred (in AdS$_4$) when $\kappa / \rho = 9/4$, i.e. when $\rho_{BF} = 3.47$. We should rescale this scale so that the instanton density fits field theory predictions - $I= 4 E^{-S_0}/L^3$ so then the model gives the BF bound violation scale as
\begin{equation}
    \rho_{BF} = 3.47 \left( \frac{4 e^{-S_0}}{L^3} \right)^{1/3}
\end{equation}
The instanton condensation scale rises to $1/L$ when $g_4^2=7.7$ - at higher $g_4$ values at the scale $1/L$ the three dimensional theory no longer has any applicability. For larger $N_c$ values $S_0\rightarrow 8 \pi^2/ g_4^2 N_c$ and eg for $N_c=3$ we find $g_4^2=5.2$.

In competition with this instanton-driven gap formation mechanism is the mechanism deduced from gap equations \cite{Miransky:1984ef,Appelquist:1986an,Cohen:1988sq,Appelquist:1996dq,Marciano:1980zf} and the study of chiral symmetry breaking in holography \cite{Jarvinen:2011qe,Kutasov:2011fr,Alvares:2012kr}. Here one follows the running anomalous dimension of the adjoint fermion bi-linear in the four dimensional theory above the scale $1/L$. One can write a holographic model here also with the dimension one  field $\hat{\lambda}$ dual to the fermion bilinear mass and condensate
\begin{equation} \label{simple}
    S_{AdS_5} = \int d^4x d \rho ~~ \rho^3 (\partial \hat{\lambda})^2 + \rho^2 {\Delta M^2} \hat{\lambda}^2
\end{equation}
Here the solutions of the equation of motion are
\begin{equation}
    \hat{\lambda} = m \rho^{-\gamma} + c \rho^{2+ \gamma}, \hspace{0.6cm} \gamma (\gamma-2) = \Delta M^2
\end{equation}

There is a BF bound violation that causes gaugino condensation when $\gamma=1$ ($\Delta M^2=-1$). One can fix the form of $\Delta M^2$ from the formula in the perturbative regime and
\begin{equation}
    \Delta M^2= - 2 \gamma \hspace{1cm} \gamma = \frac{3 C_2(R)}{2 \pi} \alpha
\end{equation}
where $C_2(R)$ is $N_c$ for the adjoint representation. This BF bound violation is now predicted to occur at $g_4^2 = 6.6$ for $N_c=2$ and
$g_4^2 = 4.4$ for $N_c=3$.

Part of our goal in this paper was to put these two mechanisms' scales in the same holographic language of BF bound violations, which we have done. In principle, one can now ask does adjoint fermion condensation or instanton condensation occur earlier? The above estimates slightly favour adjoint fermion condensation to occur first. However, one can't really deduce any such thing, since both estimates are based on wildly extrapolating the perturbative results to the non-perturbative regime. The fact that ${\cal N}=1$ super Yang-Mills theory ties the gaugino bound states and glueballs into a single multiplet suggests that the two mechanisms might merge in that theory and hence possibly all theories with adjoint matter. Holographic models are not going to resolve this fundamental question about the dynamics (first principle holographic constructions might of course).  

Thus, whether the dynamics of these theories are set at a single scale by instanton condensation or whether there are two scales, one for chiral symmetry breaking ($\gamma=1$) and one for confinement (instanton condensation) remains to be discovered. If there are two scales, one could hope to separate them. In a previous paper, we explored this phenomenon in a theory with fundamental representation and two-index symmetric representation fermion flavours \cite{Alfano:2024aek}. The idea is to let the theory trigger $\gamma=1$ for the higher dimension representation, and then run as slowly as possible (by adjusting the number of fundamentals) to a new trigger scale for fundamental representation condensation. That scale, which is in a theory similar to QCD, one posits has instanton condensation near the fundamental condensation scale. In the next section, we will repeat this method to model a theory with fundamentals and adjoint fields (neglecting instanton condensation). It is a straw-man model for the $\gamma=1$, and hence the separated scales, hypothesis which we hope will inspire first principles lattice simulations  to seek the phenomena (or disprove this world view). 

\section{Holography of QCD(Adj) + Fundamentals}

Here we present a holographic model of an SU($N_c$) gauge theory with a single Weyl fermion in the adjoint representation and in addition $N_f$ fundamental representation Dirac fermions. The model does not include the instanton sector so lives in the ``$\gamma=1$ paradigm'' - that is, we simply use the perturbative running of the anomalous dimensions for the bi-fermion operator in each representation to predict where they condense. By adjusting $N_f$, we can weaken the running between the scale where the adjoint condenses and that where the fundamental condenses to try to exhibit a gap. We assume here that instanton condensation (and hence confinement) occurs below the scale of the fundamental condensation. Likely these scales are very close as in QCD because when the fundamentals are integrated out at that condensation scale one is already at very strong coupling and the pure Yang-Mills theory in the IR will run to its pole fast.
Our model is a straw-man, intended to provoke lattice simulations to look for the gap in scales. The gaps we see depend on the extrapolations of perturbative results, and so of course come with large errors, the idea is intriguing though. The analysis mimics our previous study of these theories with two index symmetric matter rather than the adjoint where the model also predicts large gaps \cite{Alfano:2024aek}.

\subsection{The Holographic Model}

The holographic model is simply a refinement of the discussion in (\ref{simple}) above, as first presented in \cite{Alho:2013dka}. We will briefly present the action and equations of motion which will be used to calculate meson masses and decay constants for the adjoint/fundamental theories.

The gravity action for the fields for the two representations in dynamic AdS/Yang-Mills is
\begin{eqnarray} \label{eq: general_action}
S_{boson} &=& \sum_R \int d^5 x ~ \rho^3 \left( \frac{1}{r^2} (D^M X_R)^{\dagger} (D_M X_R) \right.\\
&& \left. \nonumber
+ \frac{\Delta m_R^2}{\rho^2} |X_R|^2  + \frac{1}{2 g_{R5}^2} \vphantom{\frac{1}{2}} F_{R,MN}F_{R}^{MN} + {\rm axial}\vphantom{\frac{1}{2}}  \right) \, .
\end{eqnarray}  
where the only interaction between the two representations is through their contributions to $\Delta m^2_R$. One could include cross-terms where the condensation of one representation would enhance or hinder the condensation of the other but we omit them because it is unclear how strong or what sign they should have. 

The five-dimensional coupling, as in \cite{Erlich:2005qh, Alho:2013dka}, is obtained by matching to the UV vector-vector correlator 
\begin{equation}
g_{R5}^2 = \frac{12 \pi^2}{d(R)~N_{f}(R)} \, ,
\end{equation}
where $d(R)$ is the dimension of the fermion's representation and $N_f$(R) is the number of Dirac flavours in that representation.

The model has a five-dimensional asymptotically AdS spacetime, the metric for which is
\begin{align}
ds_R^2 = r^2 dx^2_{(1,3)} + \frac{d \rho^2}{r^2} \, ,
\end{align} 
where $r^2 = \rho^2 +|X_R|^2$ - $\rho$ is the holographic radial direction corresponding to the energy scale, and with the AdS radius set to one. The $X_R$ vev is again included as a back-reaction in the metric.  

\begin{center}
\includegraphics[width=6.7cm,height=4.8cm]{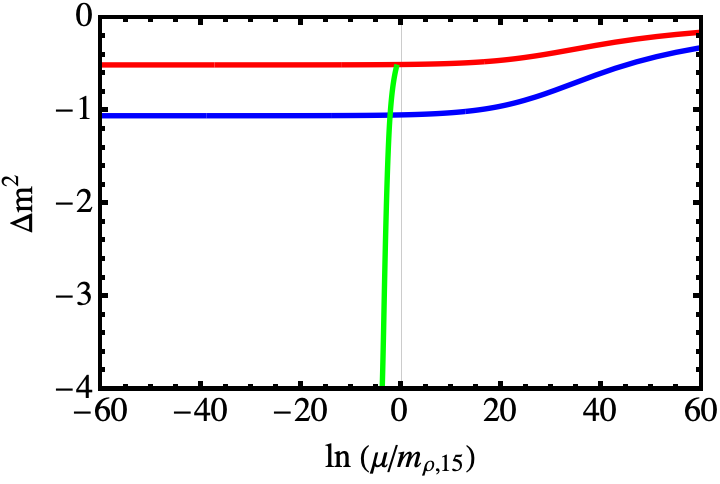}

\textit{Figure 5: SU(5) gauge theory with one Weyl adjoint field and $N^F_f=16.9$ (as an approximation to $N_f=17$): The running of $\Delta m^2$ for the adjoint rep. (blue), fundamental (red), and for the fundamentals with the higher dim rep. decoupled below the scale where it is on mass-shell leaving the fundamental running (green). The energy scales are given in units of the $\rho$-meson mass in the adjoint sector. BF bound violation occurs at $13.9$ for the adjoint and $-2.2$ for the fundamental with the adjoint decoupled.}
\end{center}

The dynamics of the particular gauge theory, with quark contributions to any running coupling, are included through $\Delta m_R^2$ in \eqref{eq: general_action}. We use the perturbative result for the running of the anomalous dimension of the quark mass, $\gamma$ and expand $M^2=\Delta(\Delta-4)$ at small $\gamma$ giving
\begin{align} \label{dm}
\Delta m^2 = - 2 \gamma.
\end{align}
Since the true running of $\gamma$ is not known non-perturbatively, we extend the perturbative results as a function of the renormalization group (RG) scale $\mu$ to the non-perturbative regime. We then directly set the field theory RG scale $\mu$ equal to the holographic RG scale $r= \sqrt{\rho^2+|X_R|^2}$.
The model breaks chiral symmetry when $\gamma$ passes 1/2, as the BF bound is then violated. This is an extrapolation of the perturbative results that favours early condensation and hence larger gaps. We explored other choices in \cite{Alfano:2024aek} but here  we seek to simply show that large gaps seem possible.

The two-loop result for the running coupling, $\alpha(\mu)$, in a gauge theory with multi-representational matter is 
\begin{align}
\mu \frac{d \alpha}{d \mu} = - b_0  \alpha^2 - b_1  \alpha^3 \, ,
\end{align} 
with 
\begin{equation}   \label{running}
\begin{array}{ccl}
b_0 &=& \frac{1}{6 \pi} \left(11 C_{2}(G) - 4 \sum_{R} T(R)N_f(R) \right) \, ,\\ &&\\
b_1 &=& \frac{1}{24 \pi^2} \left(34 C^2_{2}(G) \right. \\ &&\\ && \left.-\sum_{R} \left(20 C_{2}(G) + 12 C_{2}(R) \right) T(R) N_f(R) \right)  \, .
\end{array}
\end{equation}
where $T(R)$ is half the Dynkin index and $C_2(R)$ the quadratic Casimir (each per representation, $R$) and we have used the results for the number of Dirac fermions in a given representation (a Weyl fermion is $N_f=1/2$).

For the results to come, we have numerically  set $\alpha(1)=0.65$ which sets the strong coupling scales of the theory. We are careful though to rewrite all our results in units of the $\rho$ mass made from adjoint fields to remove this arbitrary choice. 

We use the one-loop anomalous dimension relation for the running of $\g$.
\begin{align} \label{grun}
\gamma = \frac{3~C_2(R)}{2 \pi}~\alpha.
\end{align}
We stop at one loop since it is already a guess non-perturbatively and no additional features are added beyond.

To find the vacuum of the theory  we set all fields to zero except for $|X_R| = L_R(\rho)$. For $\Delta m_R^2$ a constant, the equation of motion we obtain from \eqref{eq: general_action} is
\begin{align}
\partial_{\rho} (\rho^3 \partial_{\rho} L_R) - \rho ~ \Delta m_R^2 L_R = 0 \label{eq: vacuum qcd} \, .
\end{align}

At large $\rho$, in the UV, the asymptotic solution is $L_R(\rho) = m_R + c_R/\rho^2$, with $c_R=\brac{\Bar{q}q}$, the fermion condensates of dimension three, and $m_R$, the mass of dimension one. We numerically solve \eqref{eq: vacuum qcd} with our input $\Delta m_R^2$ for the function $L_R(\rho)$.

We use IR boundary conditions where the fermions go on mass-shell 
\begin{align} \label{vacIR}
L_R(\rho)|_{\rho=\rho_R^{IR}} = \rho_R^{IR} \, ,&& \partial_{\rho} L_R(\rho)|_{\rho=\rho_R^{IR}} = 0 \, .
\end{align}
The value of $\rho_{IR}$ is fixed in each theory. We numerically vary $\rho_{IR}$ until the value of $L_R$ at the boundary is the desired fermion mass. We refer to the vacuum solutions as $L_{R0}(\rho)$, with IR value $L_R^{IR}$ at $\rho_{IR}$.

In the models we study, the adjoint representation always condenses at a higher $\rho_R^{IR}$ than the fundamental representation. At that scale we integrate out the adjoint representation fermions and remove their contribution to the beta function at lower scales. We show an example running of $\Delta m_R^2$ for the two representations in the case of $N_c=5$ and $N_f^F=17$ in Figure 5.

The mesons of the theory are fluctuations of this vacuum configuration that satisfy the appropriate boundary conditions, matching those of the vacuum in the IR and consisting of just fluctuations of operators in the UV. The resulting Sturm-Liouville problems fix the meson masses. Decay constants obtained by substituting these solutions and those for a background source back into the action and integrating over the radial direction. The full equations can be found in \cite{Alfano:2024aek}.

\begin{center}
\includegraphics[width=6.7cm,height=4.8cm]{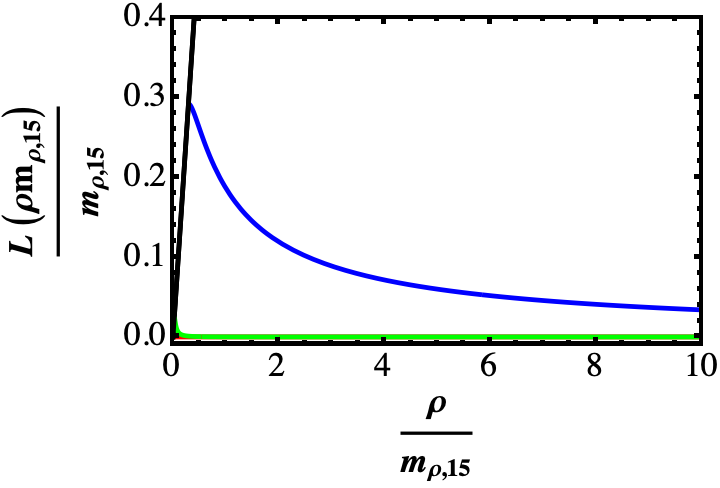}

\textit{Figure 6: SU(5) gauge theory with $N^F_f=17$ (approximates by $N_f^F=16.9$): The $L_R(\rho)$ functions for the adjoint (blue) and 5 (green) representations with $m_{IR,A}\approx0.29$ and $m_{IR,F}\approx0.025$ respectively in units of the $\rho$-meson mass in the adjoint sector.}
\end{center}

\subsection{$N_c$ and $N_f^F$ Dependence In The Massless Theory}

We will now discuss the predictions of the holographic model, starting with the case that has the largest gap in the spectrum between the adjoint and fundamental fermion bound states. 

\subsubsection{$N_c=5$ Theory}

The most extreme theory makes a good example. We take $N_c=5$ and a single Weyl fermion in the adjoint representation plus $N_f^F$ Dirac fundamental flavours.
In \cite{Evans:2020ztq} this model was identified as having a maximum gap between the condensation scales of the representations at $N_f^F=17$. There, the adjoint sector is eliminated from the running at the scale of the BF bound violation (i.e. at $\gamma_{Adj}=1$) and the $\gamma_F=1$ BF bound violation scale is then computed from the running, resulting in a separation of 22.

We implement this theory in our holographic model. We show the running of $\Delta m^2_{Adj}$ against the log of the RG scale in Figure 5. We set scales with the $\rho$-meson mass composed of adjoint fermions.
In fact, the edge of the conformal window for this theory is slightly above $N_f^F=17$, but calculations for 17 flavours are numerically intractable, 16.9 was used as a near approximation. Already at $N_f^F=16.9$, the IR fixed point value of the adjoint sector is exceedingly close to 1, with $\Delta m^2_{Adj}$ in the IR being -1.04. We can extrapolate that for $N_f^F=17$, given that the IR fixed point is even closer to 1, the violation point will be increased further and the mass gap will be slightly larger, i.e. a continuation of the trend that we will see across the range of $N_f^F$ in this section.

\begin{center}
\includegraphics[width=6.7cm,height=4.8cm]{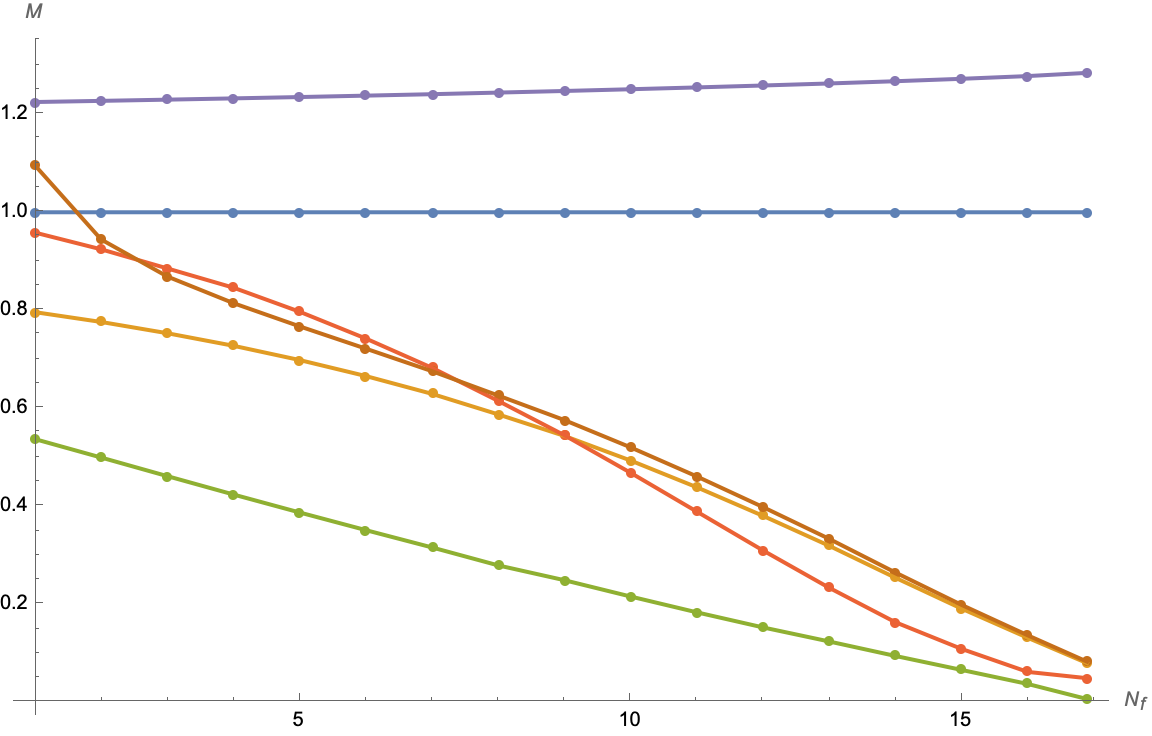}

\textit{Figure 7: Mass spectra for the SU(5) theory, $\rho$-mesons in blue (adjoint) and dark yellow (fundamental), $\sigma$-mesons in green (adjoint) and orange (fundamental), axials in purple (adjoint) and brown (fundamental). The pions in both sectors are massless at zero fermion mass.}
\end{center}

Figure 6 shows the vacuum configuration $L_{Adj}(\rho)$ in the holographic model (blue line). It corresponds to the effective fermion mass as a function of $\rho$, the RG scale. We see chiral symmetry breaking, with the line bending away from $L=0$ in the IR. The value of the lower IR value of the mass is significantly lower than the scale of BF bound violation.

Returning to Figure 5 we next consider the fundamental sector. The red and green lines show the running of $\Delta m^2_{F}$, the red being the running with the presence of the adjoint sector, and thus valid only above the scale where the adjoint goes on shell at the IR value $L_{Adj}(\rho)=\rho$. The green line then shows the results of integrating out the adjoint sector. We use an interpolation function to transition between the two runnings. BF bound violation for the fundamentals occurs at $\ln\mu=-2.2$ in units of the $\rho$ mass in the adjoint rep (compared to the value for the adjoints of $\ln\mu=13.9$).

As with the adjoint, we solve for the embedding function $L(\rho)$ for the fundamental representation, which can be seen as the green line in Figure 6. There is an IR mass gap between the two representations of 11.6. This is smaller than the factor of 22 from \cite{Evans:2020ztq} but nevertheless substantial.

We can use the holographic model to compute the spectrum and decay constants of the theory for all $N_f^F$ from 1 to 16.9 (with integer $N_f^F$ excepting the last case), beginning by computing the equivalent embedding for each value. The mass-spectrum results are shown in Figure 7. For each theory, the spectrum is normalized in terms of the adjoint representation $\rho$-meson mass, hence the blue line for the adjoint $\rho$ is a flat line at precisely 1. The decay constants are shown in Figure 8. One can see that the adjoint $\rho$-meson decay constant is also relatively unaffected by $N_f^F$.

\begin{center}
\includegraphics[width=6.7cm,height=4.8cm]{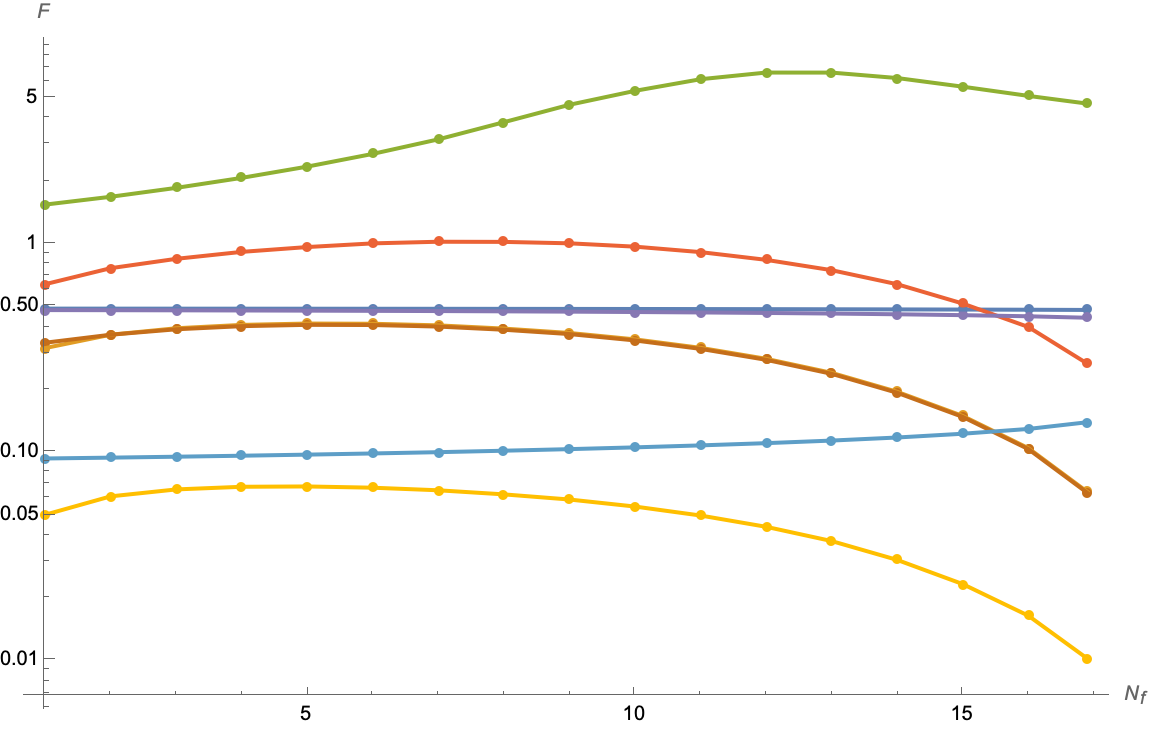}

\textit{Figure 8: Decay constants for the SU(5) theory, $\rho$-mesons in blue (adjoint) and dark yellow (fundamental), $\sigma$-mesons in green (adjoint) and orange (fundamental), axials in purple (adjoint) and brown (fundamental) and pions in cyan (adjoint) and light yellow (fundamental).}
\end{center}

For the other mesons, in the adjoint sector we see that the axial-mesons (purple) have a larger mass than the $\rho$-mesons, slightly increasing with $N_f^F$ while their decay constant is slightly smaller and decreases with $N_f^F$, though both changes are marginal.

 The adjoint sigma mesons (green) are the lightest state, reflecting the walking in the high energy theory. The mass spectrum decreasing to essentially zero by the edge of the conformal window. The adjoint sigma decay constant is the largest at all $N_f^F$, with an initial rise followed by a tail off at large $N_f^F$. Finally we have the adjoint pions which are of course, massless in the massless theory but whose decay constant (cyan) is around half that of the $\rho$-meson and only slightly affected by the increasing $N_f^F$.

In the fundamental sector, we see the $\rho$-meson (dark yellow) mass decreases strongly relative to the adjoint $\rho$ mass, with the mass gap for $N_f^F=16.9$ being $\sim12.38$. This gap should be expected to increase marginally at the actual final, non-conformal integer value of $N_f^F=17$. Meanwhile, the decay constant for the fundamental $\rho$ initially rises, before falling at large $N_f^F$ as the scale of the gap decreases. The fundamental axial-mesons begin with a larger mass than the adjoint $\rho$-meson, before falling sharply with increasing $N_f^F$, eventually dovetailing into the fundamental $\rho$-meson at high $N_f^F$, while the decay constant for the fundamental axials is almost indistinguishable from that of the fundamental $\rho$-mesons.

The fundamental $\sigma$-mesons (orange) behave similarly to their adjoint counterpart, albeit, starting at a higher mass and falling more sharply before levelling off at high $N_f^F$ as they approach the mass scale of the adjoint $\sigma$-meson. The decay constant however, along with that of the fundamental pions, displays behaviour more similar to the fundamental $\rho$ and axial-mesons, with a smooth initial rise followed by fall.

\subsubsection{Other $N_c$}

We additionally investigated the $N_c$ dependence of the spectrum, looking at the cases $N_c=2,3,4$. For each we retained the single adjoint representation Weyl fermion and $N_f^F$ fundamental Dirac fermions, computing all $N_f^F$ up to the edge of the conformal window (as per our ansatz for the running of $\gamma$). The results are shown in Figure 8. As with the SU(5) case, we computed the maximum mass gap for each theory (the ratio of the adjoint and fundamental $\rho$-meson mass spectra) which are: 

SU(2), $N_f^F=6$: 5.26;\\
SU(3), $N_f^F=10$: 10.63;\\
SU(4), $N_f^F=13$: 8.50.

\section{conclusions}

Gauge theories with fermions in the adjoint representation could potentially shed light on the origins of confinement and chiral symmetry breaking. When compactified on a small circle, such that the theories are weakly coupled, they display low energy confinement by the formation of a background density of magnetically charged instantons \cite{Unsal:2007jx}. Chiral symmetry breaking has been understood to set in when the anomalous dimension of the fermion bilinear operator becomes equal to one \cite{Miransky:1984ef,Appelquist:1986an,Cohen:1988sq,Appelquist:1996dq}. Here we have made a first step to model these two mechanisms together by presenting holographic models of both phenomena. Both the condensation of instantons and fermions are associated to BF bound violations in holographic models. A very naïve extrapolation of perturbative results suggests that the fermion condensation may occur first, but this is very far from clear cut. If the two mechanism are separable, then it is interesting to try to grow the gap between the scales. We have proposed doing this by adding additional fermions in the fundamental representation that condense at higher coupling values and slow the gauge running. A simple holographic model suggests gaps as big as an order of magnitude might be possible, although this again depends on the extrapolation of running results from the perturbative regime. Our results are intended to provoke first principle lattice simulations of such theories (in the spirit of \cite{Bergner:2020mwl}) which could shed light on the mechanism(s) of confinement and chiral symmetry breaking.

\bigskip \noindent {\bf Acknowledgements:}  
NE's work was supported by the
STFC consolidated grant  ST/X000583/1.

\newpage

\begin{center}
\includegraphics[width=6.7cm,height=4.6cm]{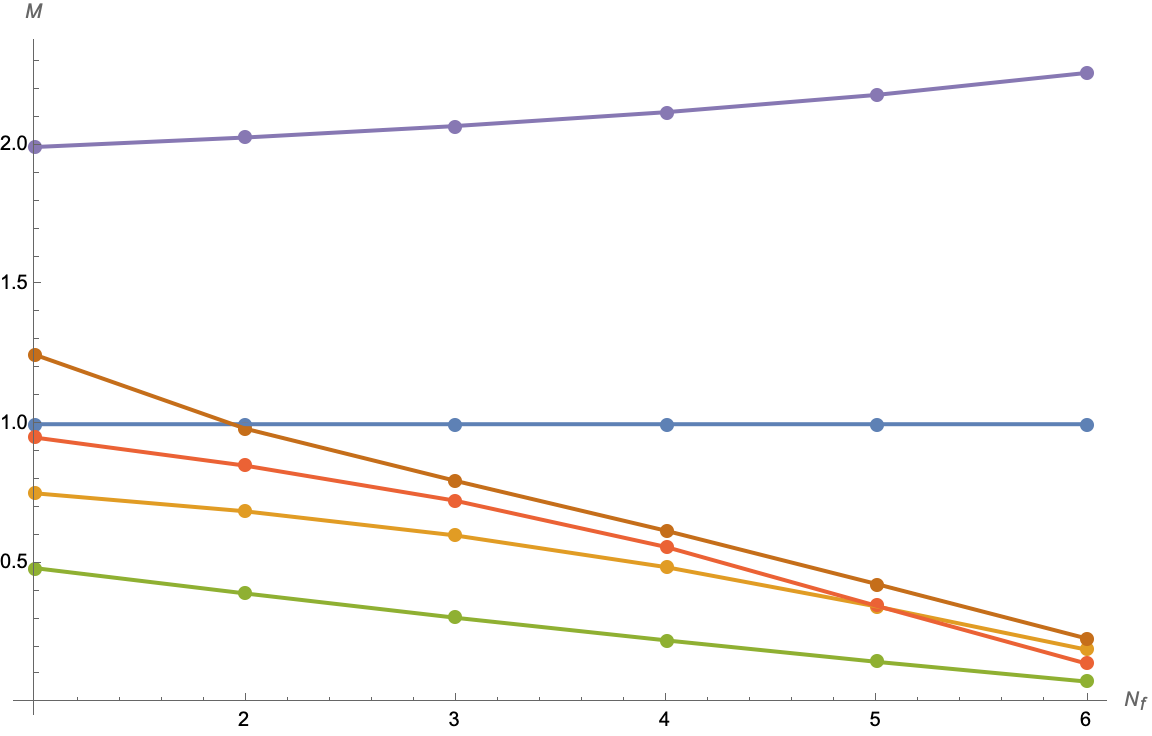}

\textit{A: Mass spectra for the SU(2) theory.}
\end{center}

\begin{center}
\includegraphics[width=6.7cm,height=4.6cm]{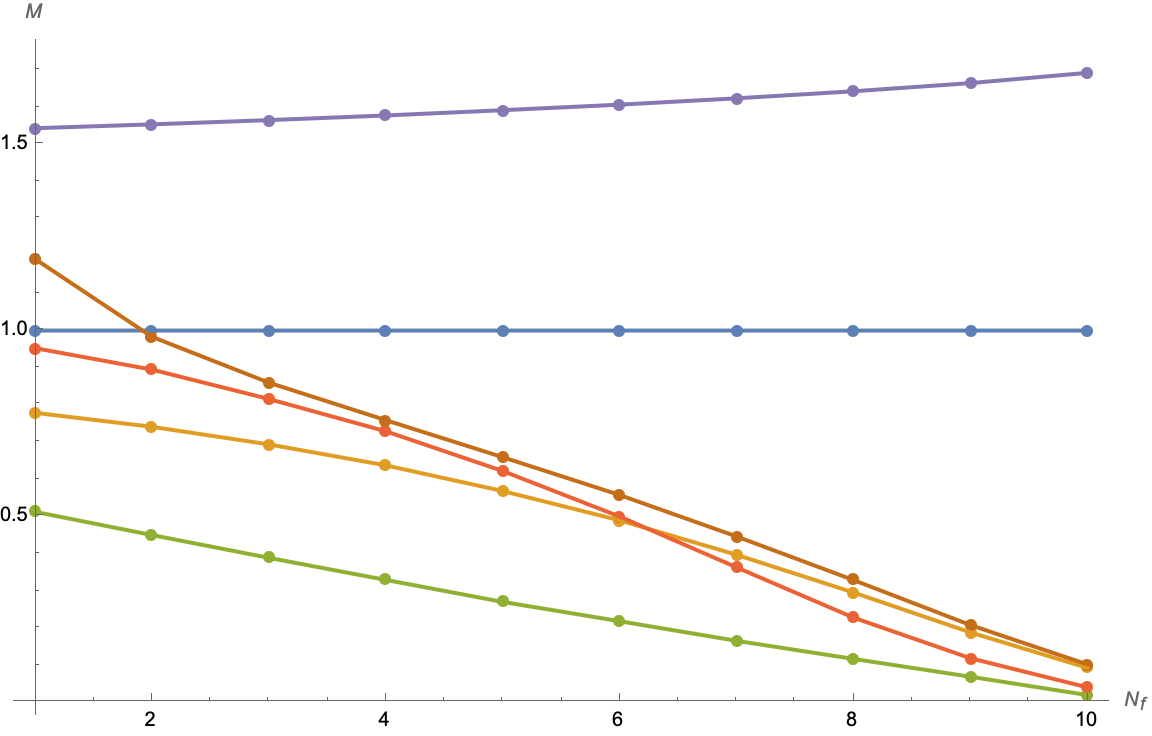}

\textit{C: Mass spectra for the SU(3) theory.}
\end{center}

\begin{center}
\includegraphics[width=6.7cm,height=4.6cm]{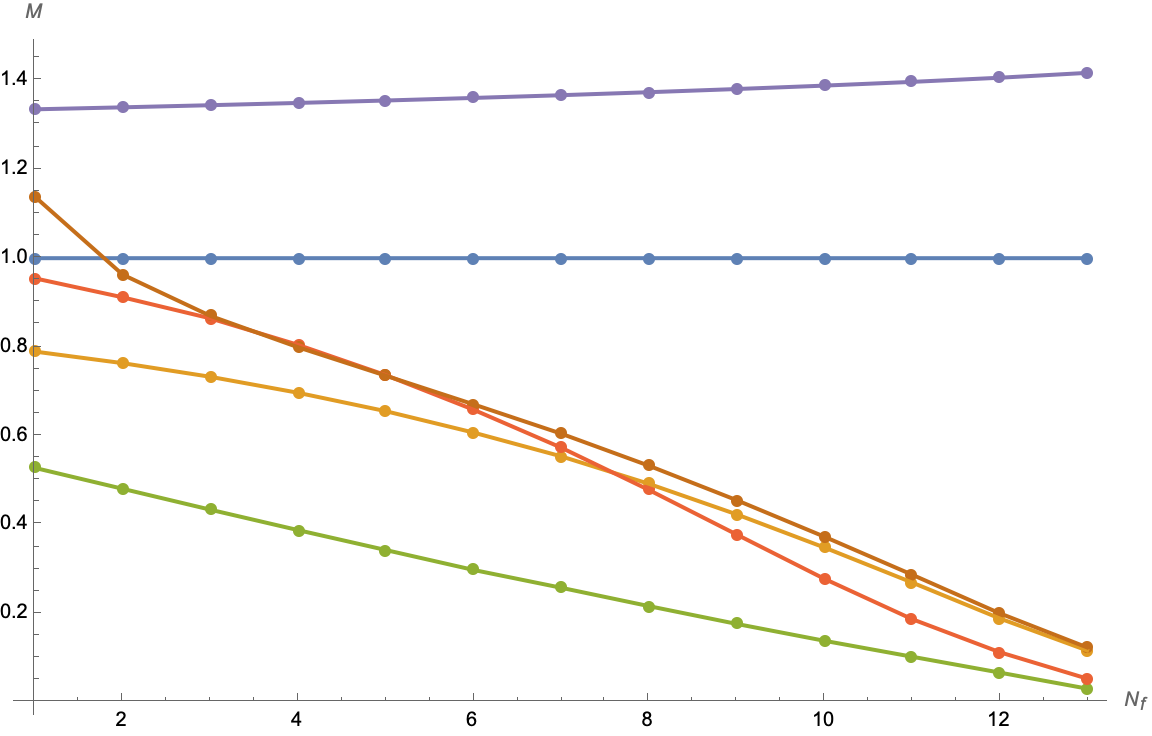}

\textit{E: Mass spectra for the SU(4) theory.}
\end{center}

\mbox{\it Figure 9: The spectra and decay constants of the SU($N_c$) gauge theory with one adjoint representation matter field } 
\mbox{\it and $N_f^F$ fundamentals for$N_c=2,3,4$. $\rho$ mesons in blue (adjoint) and dark yellow (fundamental), $\sigma$ mesons in green}
\mbox{\it 
(adjoint) and orange (fundamental), axials in purple (adjoint) and brown (fundamental) and pions in cyan (adjoint)}
\mbox{\it and light yellow (fundamental).}

$\left. \right.$ \newpage

\begin{center}
\includegraphics[width=6.7cm,height=4.6cm]{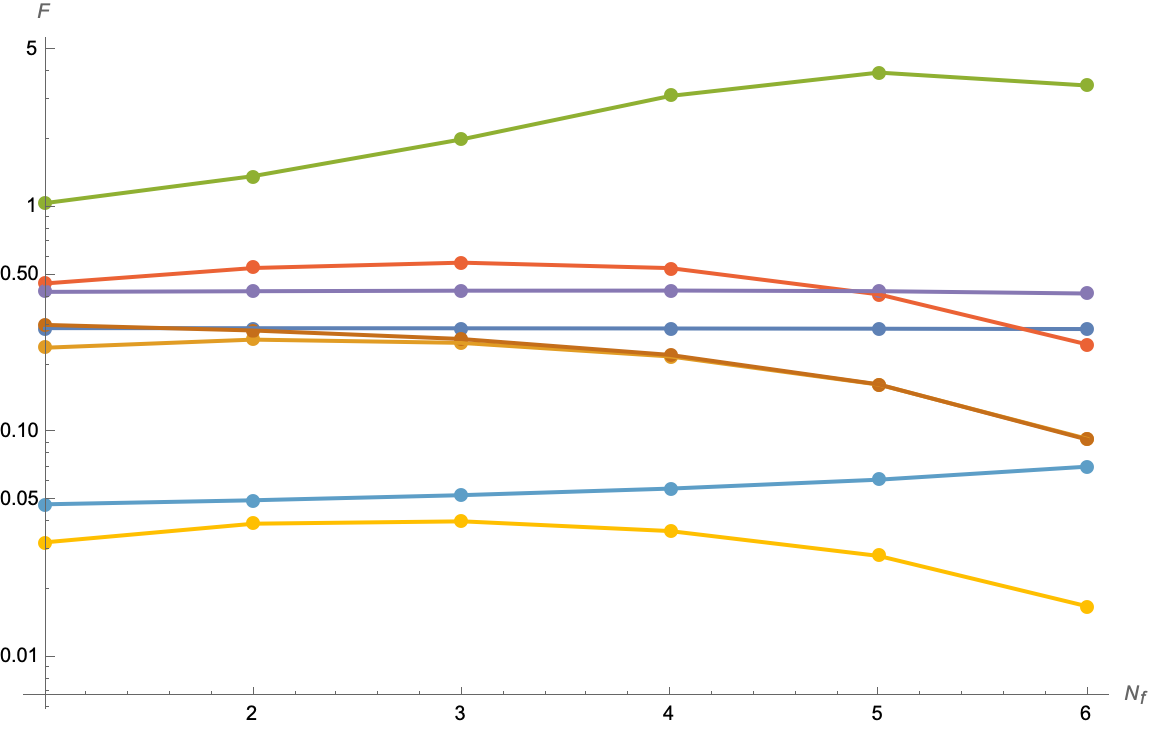}

\textit{B: Decay constants for the SU(2) theory.}
\end{center}

\begin{center}
\includegraphics[width=6.7cm,height=4.6cm]{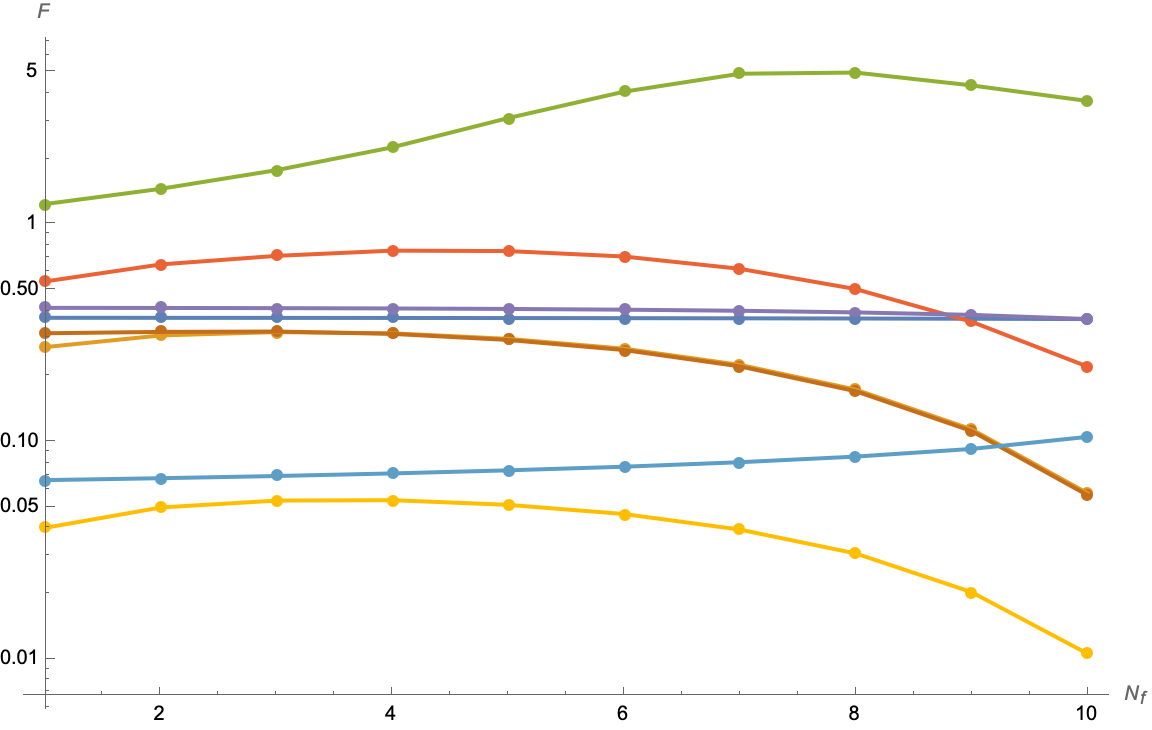}

\textit{D: Decay constants for the SU(3) theory.}
\end{center}

\begin{center}
\includegraphics[width=6.7cm,height=4.6cm]{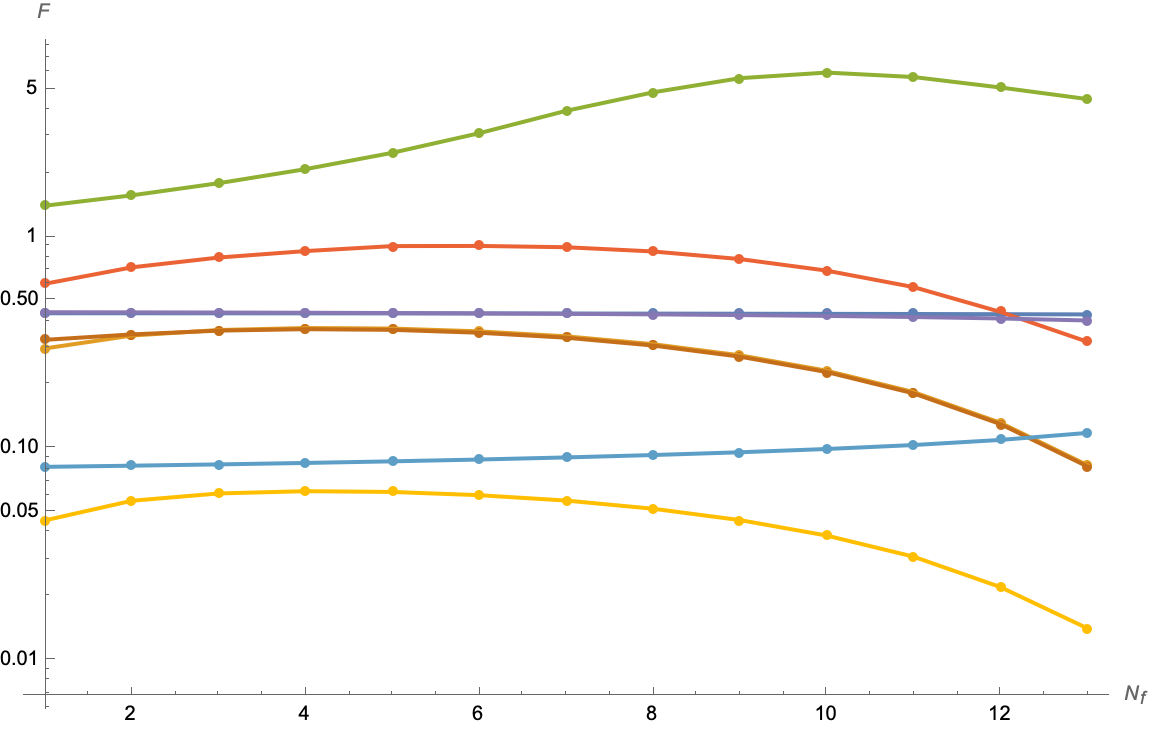}

\textit{F: Decay constants for the SU(4) theory.}
\end{center}

\newpage


\end{document}